\documentclass[12pt]{iopart}
\usepackage[dvips]{graphicx}
\usepackage{color}    

\begin{document}

\title[NO adsorption on a Cu layer]{Simulations of NO dissociative adsorption on an atomically thin Cu layer}

\author{S Nogami, H Kizaki, K Kusakabe}

\address{Graduate School of Engineering Science, Osaka University, \\
1-3 Machikaneyama-cho, Toyonaka, Osaka, 560-8531, Japan}
\ead{kabe@.mp.es.osaka-u.ac.jp}

\begin{abstract}
To investigate chemical reactivity of Cu atomic-scale structures, 
we performed simulations based on 
the generalized gradient approximation in the density functional theory. 
An atomic layer of Cu forming a triangular lattice (TL) was 
found to give a stable structure. 
The nitrogen monoxide molecule (NO) was adsorbed on 
some atomic sites of TL or on an atomic step structure (ASS) of Cu. 
The molecular adsorption energy on TL was -0.83 eV. 
Our data suggested that dissociative adsorption of NO 
with a dissociation energy of -1.08 eV 
was possible with an energy barrier of order 1.4 eV. 
In this optimized structure, 
the nitrogen and oxygen atoms were embedded in the Cu layer. 
On the step, NO adsorbed at a bridge site 
and the formation energy of Cu-(NO)-Cu 
local bond connections was estimated to be around  -1.32 eV. 
Molecular dissociation of NO 
with a dissociation energy of -0.37 eV was also possible 
around ASS. 
\end{abstract}

\maketitle

\section{Introduction}

To enhance chemical reactivity of Cu surfaces with nitrogen oxides (NO$_x$) 
is an important issue for development of new catalytic materials 
effective in the NO$_x$ reduction process \cite{Satterfield,Brown}. 
The dissociative adsorption of NO$_x$, for example, was found to be 
less expected on Cu, compared with highly reactive Rh, Ir, Ru, Co, 
and Ni surfaces, although dissociative adsorption was reported 
at finite temperatures in experiments \cite{Balkenende,Godowski}. 
In order to provide active surfaces for 
NO dissociation, Cu thin films and low index surfaces were considered 
on one hand \cite{Matloob,Johnson,So,Dumas}. 
In several electronic structure calculations based on 
the density functional theory (DFT), on the other hand, 
dissociative adsorption of NO was 
found to be possible but energetically un-favored 
compared with molecular adsorption \cite{Daelen,Gojdos,Gonzalez,Yen}. 

We note that the simulations were often performed 
with respect to reactions on stable bulk surfaces. 
Although the theoretical data suggested 
less reactivity of Cu bulk surfaces for NO$_x$ reduction, 
there could be remarkable reactivity on some surface-like 
atomic structures of Cu. 
When we considered wider classes of nano-scale structures 
other than defined surfaces of bulk Cu crystals, 
one could find another clue. In this line of approach, 
indeed, many theoretical investigations with computer simulations 
had been done intending to explore efficiency 
of {\it e.g.} step-like structures of various 
metals \cite{Loffreda,Hammer1,Hammer2,Hammer3,Hammer4,Liu,Ge,Backus,Rempel}. 

To explore possible NO dissociation, we consider ultra thin Cu structures. 
In this study, we focus on a Cu atomic layer, 
that is the triangular lattice of Cu. 
We adopted structural optimization simulations 
based on electronic structure calculations 
to find a stable Cu triangular lattice (Cu-TL). 
On this thin structure, we adsorbed an NO molecule 
and performed an optimization simulation. 
After finding molecular adsorbed structures, 
we searched possible dissociative adsorption on the Cu structures. 
To find a possible reaction path and to conclude 
a reduction process, we performed simulations for reaction path estimation. 
In the discussion of this paper, 
by comparing the obtained adsorption energies with each other, 
we will discuss a possible NO$_x$ reduction mechanism 
by using Cu nano-structures. 

\section{Methods}
\label{Methods}
We adopted the electronic structure calculation based on 
the density functional theory\cite{Hohenberg,Kohn} 
to estimate the electronic state, and 
to obtain inter-atomic forces. 
In this simulation, the Kohn-Sham wavefunctions were expanded in 
the plane-waves and the electron charge density was given 
both on a real space mesh and on the Fourier mesh. 
An approximation for the exchange-correlation energy functional 
by Perdew, Burke, and Ernzerhof\cite{Perdew2,Perdew3} 
in the scheme of the generalized gradient approximation was adopted. 
The ultra-soft pseudo-potential\cite{Vanderbilt} was 
utilized to describe the valence electron state. 
All of our simulations were done using the simulation package, 
the Quantum ESPRESSO \cite{Giannozzi}. 

The calculation conditions are summarized as follows. 
The energy cut-off for the wave function expansion was 30 [Ry], 
while the cut-off for the charge density was 240 [Ry]. 
The Brillouin zone integration was done using a $k$ mesh of 
8$\times$8$\times$1 for the largest super cell adopted. 
These values were refined, if the computation facility allowed 
much accurate calculations. 
The convergence criterion for the force constant was 
that the simulation ended, when the absolute value of 
the total force vector became less than 1$\times$10$^{-3}$ [Ry/a.u.]. 

\section{Atomic layer of Cu}
\label{Atomic_Layer}
To explore possible high reactivity of Cu nano-structures, 
we considered atomic-layer structures. 
An important structure for our discussion is 
the Cu triangular lattice (Cu-TL). 
In this section, we show data for structural and 
electronic properties of Cu-TL. 

\subsection{Atomic structure}
\label{Cu-TL}

\begin{figure}[h]
  \begin{center}
    \includegraphics[keepaspectratio=true,height=80mm]{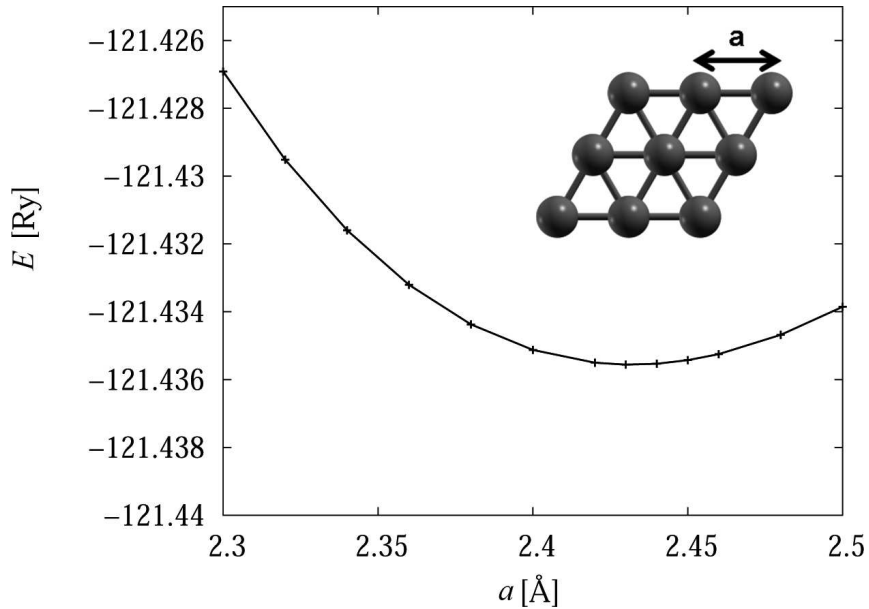}
  \end{center}
  \caption{The total energy $E$ of the triangular lattice of Cu. 
The value of $E$ in Rydberg 
is given as a function of the lattice constant $a$[\AA]. }
  \label{fig:Cu-TL-ene-a}
\end{figure}

We obtained an optimized lattice structure using 
a Cu atomic layer in a primitive super cell. 
Major calculation conditions were the same as 
those given in Section \ref{Methods}. 
The $k$-point mesh was 24$\times$24$\times$1 in this simulation. 
The cell was given in a hexagonal structure. 
The vacuum layer had thickness of 15 \AA. 
In this simulation, the value of the lattice constant was optimized. 
The bond length was found to be 2.43 \AA. 
(Fig.~\ref{fig:Cu-TL-ene-a}) 
This value is rather small compared to the bond length 2.55 \AA \
of the bulk fcc Cu. 
The reason for shrink in the bond length is mainly to reduce 
the total band energy. 
The total energy of TL was energetically higher than the bulk Cu 
by 1.2 eV per a Cu atom. 

Starting from some initial conditions, 
we found appearance of Cu-TL in optimized structures. 
As another evidence to show the local stability of Cu-TL, 
we considered an atomic two-layer structure (ATLS). 
This structure was obtained by cutting the bulk fcc Cu crystal 
and was placed in a simulation super-cell. 
The layer structure was perpendicular to the (100) direction 
of bulk Cu and thus was contained in an orthorhombic unit cell. 
Energy difference between ATLS and Cu-TL was 3.11 eV per a Cu atom. 
An optimization calculation of the structure 
concluded local stability. 
But, ATLS was not kept against global reconstruction which was 
happened when an NO molecule was adsorbed on it. 
Furthermore, we found a strongly reacted structure 
starting from an NO molecule adsorbed on ATLS. 
It means that using ATLS as an initial structure, naively speaking, 
we realized simulated annealing in our simulation. 
Compared with this un-stable structure, Cu-TL was found to be stable. 
Once the molecule was adsorbed on ATLS, reconstruction of ATLS 
happened and formation of Cu-TL was observed in our simulations. 
Conversely, we can say that Cu-TL is stable against 
distortion making corrugation toward ATLS. 

Even when one observed local stability of an atomic structure 
in simulation, however, a final evidence of 
the structure would be requested to be given using real experiments. 
Realization of an atomically thin layer, 
{\it i.e.} Cu-TL, will need development of a fabrication method. 
Recently, formation of an atomic layer of Pb 
on the Si(111) surface was reported \cite{Pb-layer}. 
In this superconducting Pb system, positions of Pb atoms are affected by 
the atomic structure of the substrate and inter-atomic distance 
between Pb atoms is not determined independently from the substrate. 
The most remarkable example of natural realization of the atomic layer 
is graphene \cite{Graphene1,Graphene2}. 
This unique flexible structure of carbon is possible to be 
supported in air according to the strong C-C sp$^2$ bonding. 
Peeling a graphene sheet and pasting it on a silicon-di-oxide surface, 
graphene is obtained efficiently from graphite. 

In case of Cu, we may expect formation of an atomic layer 
on a suitable inert substrate. We might be able to keep 
the atomic layer as a film pasted on a support 
with a nano-meter-scale hole. 
Then, mechanical properties of the atomic layer 
would be paid attention like a graphene sheet \cite{Mohr}.
But for our consideration of NO adsorption, 
a local structure of Cu is important. 
So, we assume an atomic-scale local structure 
in a part of nano-scale Cu. 

\subsection{Electronic structure}
\begin{figure}[h]
  \begin{center}
    \includegraphics[keepaspectratio=true,height=80mm]{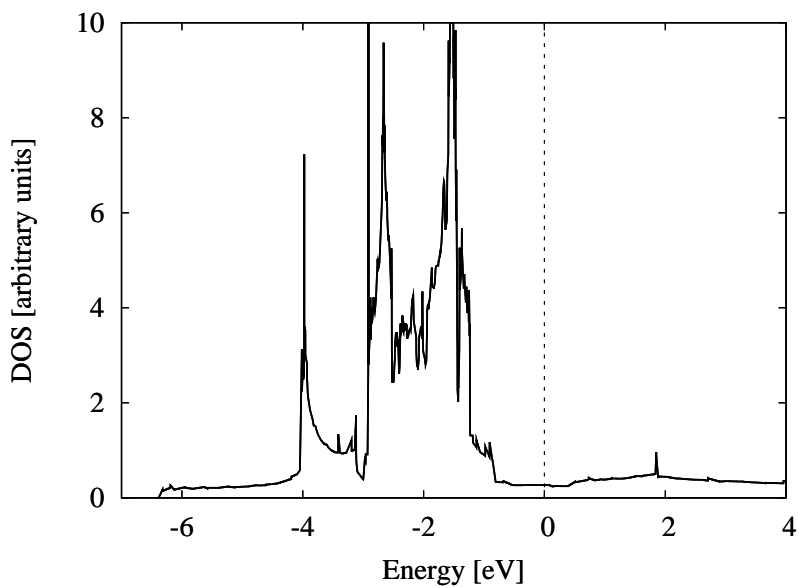}
  \end{center}
  \caption{The electron density of states of the triangular lattice of Cu.}
  \label{fig:cu111-1-tetra}
\end{figure}

\begin{figure}[h]
  \begin{center}
   \includegraphics[keepaspectratio=true,height=70mm]{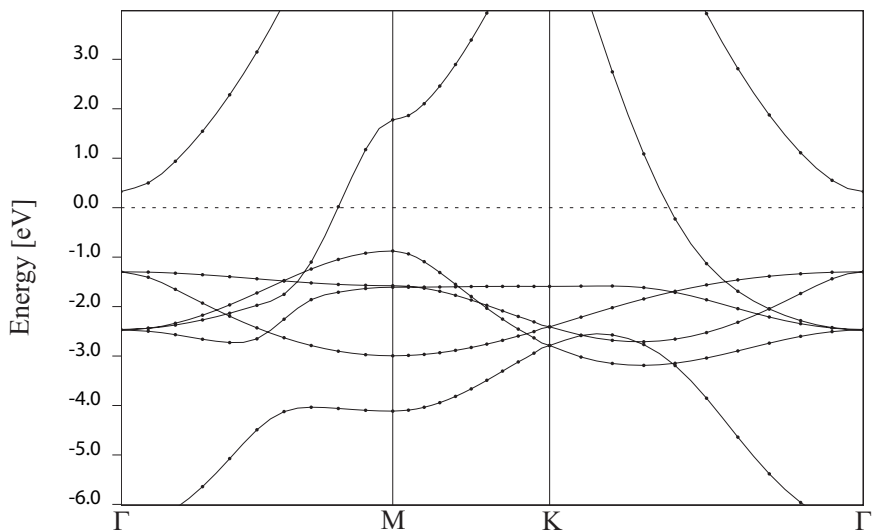}
  \end{center}
  \caption{The Kohn-Sham band structure of the Cu triangular lattice.}
  \label{fig:cu-2d-band}
\end{figure}

The density of states (DOS) of Cu-TL is shown in Fig. \ref{fig:cu111-1-tetra}. 
The major peaks are characterized similarly to the bulk copper. 
Looking DOS from the low energy region, we see that 
the 4s band starts from -6.38 eV and spreads over above 
the Fermi energy. 
Sharp peaks of 3d levels are seen from -4.04 eV but the 3d 
bands end below the Fermi level. 
Thus, parfectly filled 3d band with 
the $(3d)^{10}$ configuration is kept 
and the structure behaves as an $s$ metal. 
These characteristic features are seen in 
the electronic band structure of Cu-TL, too. (Fig. \ref{fig:cu-2d-band}) 
Along the $\Gamma$-M line, or 
the K-$\Gamma$ line, we see hybridization of 
the 4s band and a 3d band. 

Comparison with the Cu (111) surface 
allows us to evaluate 
similarity and difference between Cu-TL and the bulk Cu surface. 
The density of states in the 4s band leveled almost around 
0.2 owing to two-dimensional nature. 
DOS in the 3d bands peaked well above 10 for Cu-TL 
in the unit of states/eV per a unit cell, 
while the value is from 4 to 6, except for 
a singularity at the top of the 3d bands, for the Cu (111) surface. 
The shorter bond length of Cu atoms, the height in DOS 
should be lower in a fixed lattice structure. 
Thus, the higher DOS peak for Cu-TL than 
for the bulk Cu suggests that 
two-dimensional nature of Cu-TL affects DOS. 

Since DOS at the Fermi level 
is almost the same for both Cu-TL and the Cu (111) surface, 
the chemical reactivity of NO is expected to be similar, 
if the structure is kept undeformed. However, 
we should note that the top of Cu 3d bands is much closer 
to the Fermi energy for Cu-TL than the Cu (111) surface. 
This tendency suggests higher reactivity of Cu-TL against NO. 

When the stable adsorption site is the on-top site, 
similarity in characteristic energy like the adsorption energy 
would be expected. 
However, if a bridge site or a hollow site became stable for NO 
on Cu-TL, we could have difference even in the chemical reactivity 
from that on the Cu (111) surface. 
This is because much easy deformation of the Cu network structure 
is expected for Cu-TL and the bond formation between NO and Cu-TL 
will create distortion. 
In the next section, we discuss occurrence of strong reactions 
between NO and Cu-TL. 

\section{Adsorption of NO}
We consider adsorption of NO on Cu-TL and 
an atomic step like structure (ASS) created on an Cu atomic layer. 
The second structure was found in optimization 
simulations of NO adsorption on ATLS. 
Observing results of NO-adsorbed structures starting from ATLS, 
we identified a stable substrate structure 
in a super cell as Cu-TL with ASS in our simulations. 
Therefore, we regarded Cu-TL and ASS as typical 
atomic-scale layer structures of Cu. 
Several characteristic adsorption sites for NO were 
found on these structures. 

\subsection{Adsorbed structures}
\label{Ads-Structure}
Molecular adsorbed structures were obtained 
by structural optimization. 
Starting from an initial structure 
with a NO molecule a little separated from a substrate, 
Cu-TL, ASS or ATLS, each adsorbed structure was determined. 
By a series of simulations, we found the next general rules for 
molecular adsorption. 
On Cu-TL, adsorption on a hollow site is energetically most favorable. 
On ASS, a bridge site on the Cu array is energetically most favorable 
among sites including 
an on-top site, a hollow site in the back surface, and a bridge site 
in the back surface. 
Thus, we treat these locally stable structures only in the following discussion. 

\begin{figure}[h]
  \begin{center}
    \includegraphics[keepaspectratio=true,height=80mm]{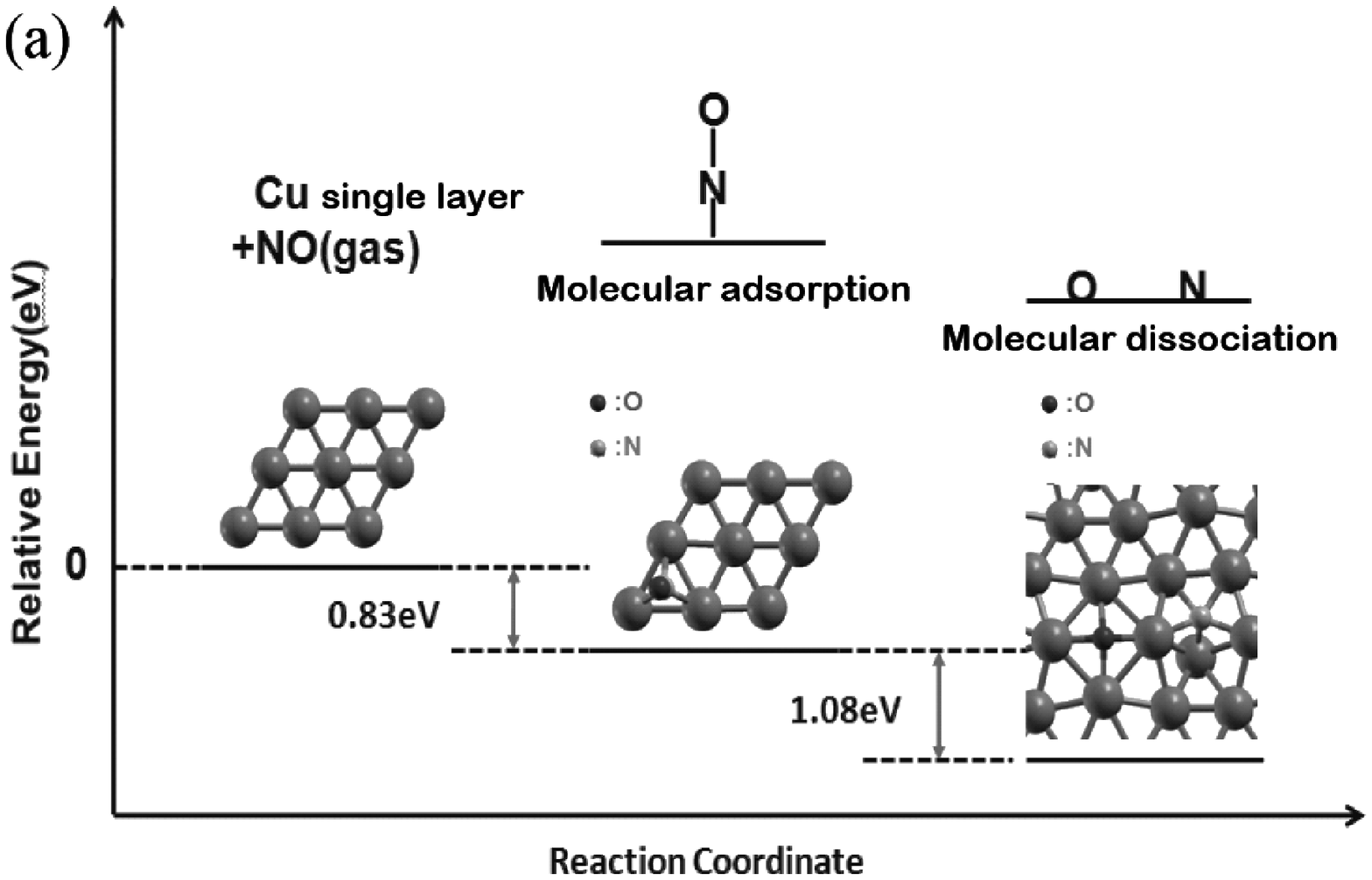}
  \end{center}
  \begin{center}
    \includegraphics[keepaspectratio=true,height=80mm]{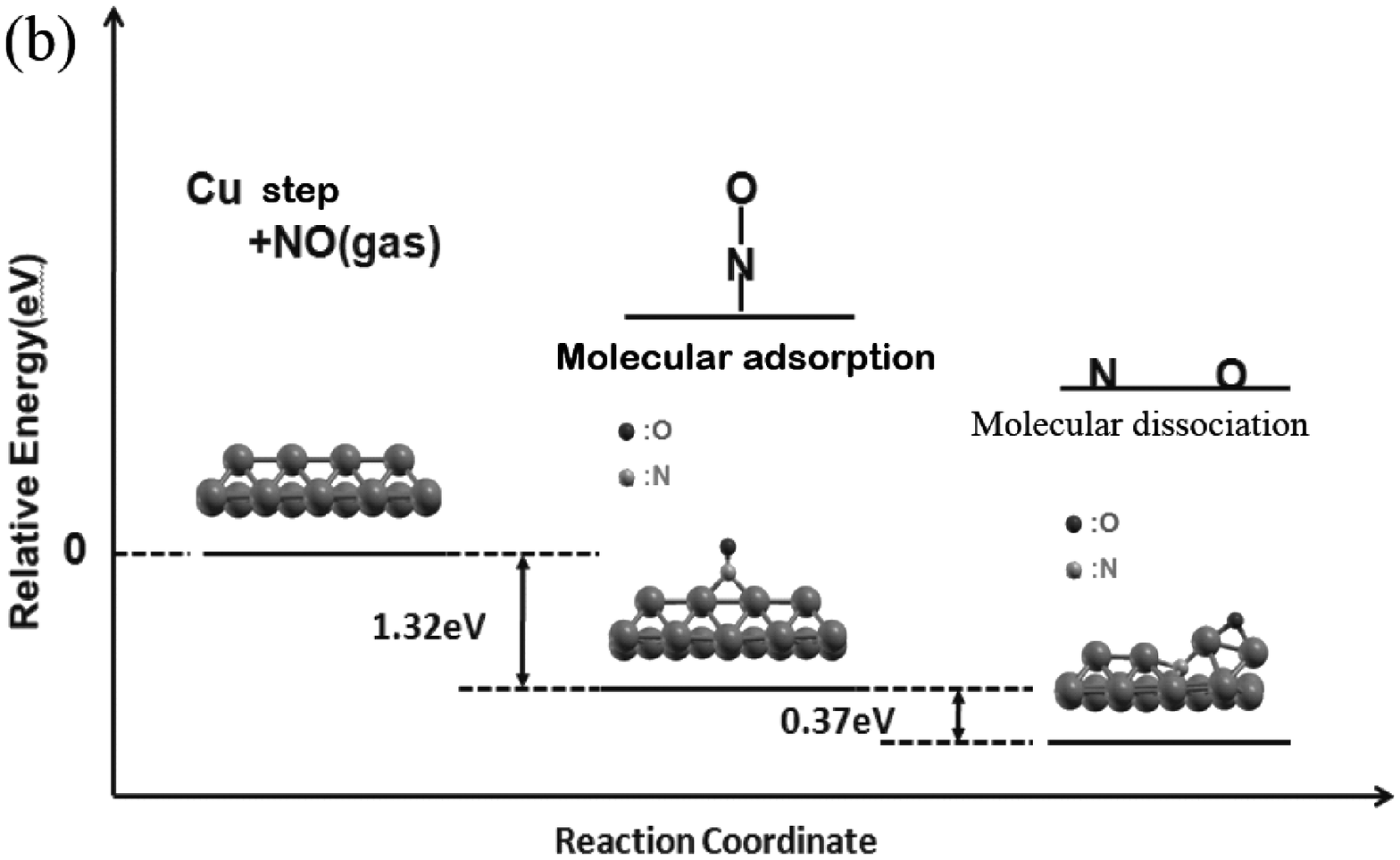}
  \end{center}
  \caption{The energy diagram of NO adsorbed structures on a Cu atomic layer in
(a) the triangular lattice and in (b) a layer with an atomic step-like structure.}
  \label{fig:ss-sideview}
\end{figure}

While, structures corresponding to dissociative adsorption 
were given by locating N and O atoms a little separated on the substrate 
and by optimizing the whole structure. 
We have found two locally stable dissociative adsorbed structures 
on Cu-TL and on ASS. 
The structures are depicted in Fig.~\ref{fig:ss-sideview}. 
As typical structures, we consider these structures only. 

We define the adsorption energy by the next formula. 
\begin{eqnarray}
E_{ad:NO}&=&E_{Cu-NO}-E_{Cu}-E_{NO}, \\
E_{ad:N,O}&=&E_{Cu-N,O}-E_{Cu}-E_{NO}.
\end{eqnarray}
Here, $E_{ad:NO}$ is the molecular adsorption energy, 
while $E_{ad:N,O}$ is the dissociative adsorption energy. 
The values of $E_{Cu-NO}$ and 
$E_{Cu-N,O}$ are the total energy of 
a Cu slab with NO and that of another slab with 
a N atom and an O atom adsorbed on the Cu slab, 
$E_{Cu}$ is the total energy 
of a Cu slab without NO, and $E_{NO}$ is the total energy of 
the NO molecule contained in a super cell with the same size as 
the other calculations. Molecular dissociation energy is defined as, 
\begin{eqnarray}
E_{diss}&=&E_{ad:N,O}-E_{ad:NO}. 
\end{eqnarray}

Adsorbed structures found in our simulations are 
itemized in the next list. 
The adsorption energy is also shown 
in each parenthesis for convenience. 

\begin{description}
\item[molecular adsorption on Cu-TL] 
In adsorption of a NO molecule on a surface of Cu-TL, 
a hollow site (-0.83 eV) is selected. 
See the center figure of Fig.~\ref{fig:ss-sideview} (a). 

\item[molecular adsorption on ASS] 
In adsorption of a NO molecule on an atomic step-like structure, 
a bridge site (-1.32 eV) is selected. 
See the center figure of Fig.~\ref{fig:ss-sideview} (b). 

\item[dissociative adsorption] 
Dissociative adsorption of NO is found on TL 
(-1.92 eV) and on ASS (-1.69 eV). 
See the right figures of Fig.~\ref{fig:ss-sideview} (a) and (b). 
\end{description}

Now, dissociative adsorption structures are discussed. 
We have two typical dissociative adsorption structures 
on Cu-TL and on ASS. 
In the first structure, the nitrogen atom locates at 
a center of five surrounding Cu atoms. 
(See the right figures of Fig.~\ref{fig:ss-sideview} (a).)
This structure may be regarded as 
a nitrogen interstitial impurity in a Cu 
lattice. 
Arrangement of Cu is largely distorted from pure TL, so that 
the N atom is embedded in Cu layer.
The oxygen atom locates at a hollow site and 
it is embedded in Cu layer. 
The local structure of these impurity sites are 
NCu$_5$ and OCu$_4$. 
Here, the OCu$_4$ structure is planer. 

A reason for appearance of 
the high coordination numbers for N and O 
is that Cu valence is not largely modified and that 
3d$^{10}$ configuration is almost kept. 
4s electrons are in extended states so that the local N$^{-3}$ 
and O$^{-2}$ are efficiently screened by 
neighboring five and four copper atoms. 

On SS, we have another dissociative 
adsorbed structure for NO. 
The nitrogen atom is again embedded in the Cu structure. 
(See the right figure of Fig.~\ref{fig:ss-sideview} (b).) 
In this structure, the oxygen atom is at a bridge site and 
keeps two-fold coordination, 
while the nitrogen atom has four-fold coordination. 

Now we compare the obtained values of adsorption energy 
with those in the literature. 
The molecular adsorption energy is higher for ASS than those on Cu-TL. 
Except for a case of a bond-center site on the back surface of ASS, 
molecular adsorption favors the bridge site of the step-like structure. 
This general tendency is natural in comparison to the other 
examples known in the literature. 

The most notable feature of our results is 
the finding of the large dissociative adsorption energy. 
We conclude that the dissociative adsorption may happen, 
when the nitrogen atom can go to an interstitial site of 
a Cu structure. The large value of $E_{ad:N,O}$ 
(-1.92 eV on TL and -1.69 eV on ASS) 
is actually possible, since these structures 
possess high coordination of Cu around the nitrogen atom. 

On the clean Cu (111) surface, NO favors 
the molecular adsorption (with the adsorption energy of 
-1.22 eV estimated in Ref. \cite{Gojdos}) 
against the dissociation of NO, 
where the dissociative adsorption energy is estimated to be 
-0.79 eV in Ref. \cite{Gojdos}. 
The qualitative difference between data for known bulk surfaces and our result 
should be attributed on movement of Cu atoms in the reconstruction process. 
In our simulation, positions of Cu atoms are rather easily modified 
because the atomic structure of Cu is just a single layer. 
Even in the optimization simulation, we can reach the nitrogen insertion 
in the Cu structures. 

From the present result, we conjecture the following picture. 
If a Cu structure allows large configurational distortion 
owing to the chemical reaction with NO, 
the nitrogen atom can move into the Cu structure and form 
the local NCu$_m$ configuration. ($m=4$ or 5.) 
Owing to the energy reduction coming from the large formation energy 
of the local NCu$_m$ structure, 
we can expect even the dissociative adsorption of NO on Cu. 
In the real Cu nano-structures, there can happen large distortion of 
Cu configuration owing to finite temperatures and possible local strain. 
Therefore, our simulation, which is prepared using the atomic layer 
of Cu, might have derived a hidden possible path of 
the NO dissociative adsorption on Cu structures. 

\subsection{Reaction path estimation of molecular dissociation}
To estimate a reaction path on the NO reduction on Cu-TL, 
we estimated the local structure and the energy 
of a transition state using the nudged-elastic-band method. 
The initial configuration was the molecular adsorbed structure on Cu-TL 
and the final configuration was the dissociative adsorbed structure 
obtained in \S \ref{Ads-Structure}. 
The dissociation reaction was determined by obtaining 
the transition state with an energy $E_{TS}$ 
for the NO reduction process. 

The activation energy for 
dissociation of NO on the atomic layer is estimated 
using the next definition. 
\begin{eqnarray}
\Delta E_{diss}&=&E_{Cu-NO}-E_{TS}. 
\end{eqnarray}
An upper bound of the transition potential 
barrier is estimated to be 1.4 eV. 
In the initial state, the NO molecule adsorbed on Cu-TL with 
the nitrogen atom binding to the Cu surface. 
In the transition state, the oxygen atom 
had local bond connections with surrounding Cu atoms to 
reduce the total energy. 
To form this distorted structure in the transition state, 
the whole Cu atomic configuration were optimized, 
creating drastic change in Cu-TL. 

\section{Summary and Conclusion}

Utilizing the DFT-GGA simulations, 
we have shown that NO dissociative adsorption may happen 
on an atomic Cu layer, which is the triangular lattice of Cu atoms (Cu-TL). 
The reactivity of Cu-TL against molecular adsorption of NO 
was found to be similar to the Cu(111) surface. 
Some stable sites for NO were found to give molecular adsorption. 
However, our optimization simulation revealed that there was 
a co-adsorbed structure of N and O atoms, 
which was energetically stabler by 1.08eV 
than the molecular adsorbed Cu-TL structure. 
A reaction path estimation showed existence of 
a path with an energy barrier of 1.4eV. 
Thus, we may conclude molecular dissociation of NO on the Cu atomic layer. 
The large dissociation energy appears 
owing to formation of local N-Cu or O-Cu bondings and  
creation of local N-Cu$_n$ and O-Cu$_m$ structures. 

We further considered an atomic step-like structure (ASS) of Cu, 
which was an atomic-scale wrinkle in the Cu-TL structure. 
The absolute value of the molecular adsorption energy 
on the step was larger than the values found for Cu-bulk surfaces or Cu-TL. 
Our simulation revealed that 
there existed a dissociative adsorbed structure 
in which a nitrogen impurity site embedded in a Cu structure was created. 
The estimated dissociation energy of NO became -0.37 eV on ASS. 

Flexibility against modification of this Cu atomic structure 
in the nano-meter scale is decisive both to stabilize 
dissociative adsorption of NO 
and to reduce the energy barrier on the NO-reduction path. 
Catalytic activity of Cu to reduce NO should 
appear on the atomically flexible Cu networks. 
Therefore, in order to realize Cu-based NO$_x$ reduction catalysts, 
it is important to create atomic structures of Cu, {\it i.e.} 
atomic layers, atomic scale clusters, 
and atomic scale networks, 
which allow conformational change. 

\subsection*{Acknowledgement}
This work was supported by 
the elements science and technology project 
and also by Grant-in-Aid for Scientific Research in 
Priority Area (No. 19051016) and 
a Grants-in-Aid for Scientific Research (No. 22360049). 
The computation is partly done using the computer facility of 
ISSP, Univ. of Tokyo, and RI2T, Kyushu University.

\end{document}